\documentclass[aps,prl,twocolumn,a4paper,floatfix,showpacs]{revtex4-1}
\usepackage{hyperref}
\usepackage{natbib}
\usepackage{graphicx}
\usepackage{bm}
\usepackage{epstopdf}
\usepackage{color}
\usepackage{amsmath}
\usepackage[capitalise]{cleveref}

\usepackage{mathtools}
\newcommand{\defeq}{\vcentcolon=}

\newcommand{\vx}{\bm{x}}
\newcommand{\vu}{\bm{u}}
\newcommand{\vU}{\bm{U}}

\newcommand{\dx}{\partial_x}
\newcommand{\inty}{\int_{-1}^1 dy}
\newcommand{\diff}[2]{\frac{d #2 }{d #1 } }
\newcommand{\diffp}[2]{\frac{\partial #2 }{\partial #1 } }

\begin{document}
\title{Chaotic self-sustaining structure embeded in turbulent-laminar interface}
\author{Toshiki Teramura}
\email{teramura@kyoryu.scphys.kyoto-u.ac.jp}
\affiliation{Department of Physics and Astronomy, Graduate School of Science, Kyoto University, Japan}
\author{Sadayoshi Toh}
\affiliation{Department of Physics and Astronomy, Graduate School of Science, Kyoto University, Japan}
\date{\today}
\begin{abstract}
An iterface structure between turbulence and laminar flow is
investigated in two-dimensional channel flow.
This spatially localized structure not only sustains itself,
but also converts laminar state into turbulence actively.
In other words, this coherent structure has a functionality
to generate inhomogeneity by its inner dynamics.
The dynamics of this functional coherent structure is isolated
using the filtered simulation,
and a physical perspective of its dynamics is summarized
in a phenomenological model called an ``ejection-jet'' cycle,
which includes multiscale interaction process.
\end{abstract}
\maketitle
\section{Introduction}
Turbulence ubiquitously appears in nature:
from qaurk-gluon plasma \cite{Venugopalan2014}
to the Universe \cite{Balbus1998}.
Because of its strong nonlinearity, most studies related to
turbulence may have adopted more or less
statistical or coarse graining approaches
\cite{Frisch1995}.
Though they have vividly revealed phenomenological and/or kinematic natures of turbulence
such as the energy transfer among different scales and places,
these statistical treatments are not sufficiently adequate
to elucidate concrete mechanisms of even such fundamental
processes of turbulence:
For example, what substance, e.g. vortices, transfers energy
or why the energy transfer occurs.
On the other hand, the dynamical systems approaches to turbulence
have helped us describe these mechanisms
with numerically obtained components (invariant sets)
in the phase space such as fixed points, periodic orbits and
their connections \cite{Kawahara2012}.

Recent developments in the dynamical systems approach
to turbulence arrive at the next stage,
where the spatial inhomogeneity is taken into account.
Famous actors on the previous stage are the ``minimal'' flows
\cite{Jimenez1991},
which mean direct numerical simulations with
minimal system sizes reproducing elementary processes
and some statistical quantities of turbulence.
The phase spaces embeding them are effectively low-dimensional.
However, those of the spatially inhomogeneous turbulent flows
are no longer low-dimensional,
and it is quite hard to treat such high-dimensional phase spaces
both theoretically and numerically.

One simple strategy to overcome this high-dimensionality is
to consider spatially localized self-sustaining structures
as building blocks (BBs) of turbulence.
Though each of BBs may consist of internal fundamental elements, 
each block is expected to be effectively low-dimensional like the minimal flows.
Indeed, various types of numerical exact solutions
to the Navier-Stokes equation corresponding
to localized coherent structures have been obtained so far
in pipe flow \cite{Avila2013},
plane Couette flow \cite{CambridgeJournals:7324528,Schneider2010a,Eckhardt2014a}
and asymptotic suction boundary layer \cite{Khapko2013}.
Then the dynamics of spatially extended systems are expected to be decomposed
into that of each localized coherent structure and their interactions.

At first glance
this building block strategy may be incompatible with global inhomogeneity
since they are introduced to represent local dynamics.
One scenario to treat global inhomogeneity in this strategy
is to regard it as a collective dynamics among BBs.
Since each of BBs is represented by a low-dimensional model,
a coarse-grained model governing their interactions can be deduced
as done for the chemical oscillations \cite{Kuramoto2003}.
This phase reduction \cite{Kuramoto2003} scenario has succeeded
in explaining properties of ``puff'' in pipe flow \cite{Barkley2011}.
We demonstrate, however,
that this scenario breaks down at least for a fundamental inhomogeneous system,
namely turbulent-laminar interface.

Instead, we try to deal with global inhomogeneity
by extending roles of each BB.
We focus on a turbulent-laminar interface in two-dimensional channel flow.
As clarified in this paper,
a localized self-sustaining structure is embedded in the interface.
This structure, which we call chaotic interface (CI),
produces turbulence downstream by its inner dynamics
while invading upstream laminar flow.
The turbulent-laminar interface is governed by CI in this sense;
i.e. this global inhomogeneity is generated by the local dynamics.
We introduce the term ``functional'' coherent structures (FCS)
to represent such active localized coherent structures.
We clarify the dyanamics of CI in detail
and evaluate how this functional building block scenario
explains this global inhomogeneity.

\section{chaotic interface structure}
\begin{figure*}
 \centering
 \includegraphics[width=\textwidth]{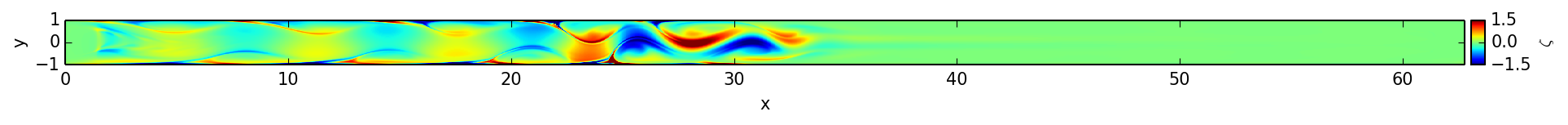}
 \caption{
  A snapshot of turbulent vorticity field.
  $\zeta$ varies from $-25$ to $25$ on the walls,
  and regions $\zeta > 1.5$ ($\zeta < -1.5$) are colored
  by the same color of $\zeta = 1.5$ ($\zeta = -1.5$).
 }
 \label{fig:Zeta}
\end{figure*}
Two-dimensional laminar channel flow has the same critical Reynolds number $Re_c$
as three-dimensional one.
In 2D case, the TS-wave solution appearing at this critical point
bifurcates into a weak chaotic state, which we call chaotic TS-wave,
as its Reynolds number increases
\cite{Jimenez1990, Fortin1994a, Umeki1994, Rauh1995}. 
In this paper, we consider a channel which contains
turbulent downstream region and laminar upstream region.

We adopt a frame of reference moving at a speed $c_I$
against the laboratory frame for CI not to march.
The streamwise and the wall-normal coordinates are denoted 
by $x$ and $y$, respectively in this interface frame.
The system is non-dimensionalized by the half width of 
the channel, so $y \in [-1,1]$.
$\vu$ denotes the velocity field in this frame.
We deal with a very long box $[0, 20\pi] \times [-1,1]$ periodic in $x$,
to emulate the dynamics realized in an infinitely long channel.
Since the walls move in the interface frame,
the non-slip boundary conditions become $\vu(x, \pm 1) = -c_I \hat\vx$,
where $\hat\vx$ denotes the $x$ directional unit vector.
The Raynolds number $Re$ is fixed to $8000$ in this paper.

To analyze the dynamics of this process in a finite computational box,
we have to keep supplying laminar region
since the turbulent region becomes wider as time goes on.
We resolve this problem using the damping filter \cite{Teramura2014}
in the interface frame.
We introduce a linear damping term
into the incompressible Navier-Stokes (NS) equation
to reproduce a laminar Poiseuille flow $\vU_L = (1-y^2-c_I)\hat\vx$
in a small region $\Omega = [0,1.4] \times [-1,1]$:
\begin{align*}
 \diffp{t}{\vu} + \left( \vu \cdot \nabla \right) \vu
 &= - \nabla p + \frac{1}{Re} \nabla^2 \vu - H_{\sigma^2, \Omega}(x) \left(\vu - \vU_L \right), \\
 H_{\sigma^2, \Omega}(x) &= \frac{1}{\sqrt{2\pi \sigma^2}} \int_\Omega dx^\prime \exp\left( \frac{(x-x^\prime)^2 }{ 2\sigma^2} \right),
\end{align*}
where the last term of NS equation is the damping filter term.
Since $c_I = 0.855$ is faster than the phase velocity of the chaotic TS-wave,
this damping term laminarizes it,
and the laminarized flow returns upstream due to the periodic boundary condition.

In this setting a turbulent-laminar interface is simulated permanently.
A snapshot is displayed in \cref{fig:Zeta} using the turbulent vorticity
$\zeta = (\nabla \times (\vu - \vU_L))_z$.
This figure shows that there are three regions:
weak turbulence ($x \lesssim 20$),
chaotic interface ($20 \lesssim x \lesssim 34$),
and laminar ($34 \lesssim x$) regions.
Moreover, the chaotic interface contains dynamic inner structures.
The snapshot shows a meandering bulk structure
and strong wall shear layers.
The weak turbulence consists of spatially modulated chaotic TS-waves.
The chaotic interface is nothing but FCS,
and we will reveal in the following that it generates the weak turbulence.
We first investigate the energy balance of these regions,
and then construct a phenomenology
for its self-sustaining mechanism and functionality.

\begin{figure}
 \centering
 \includegraphics[width=\columnwidth]{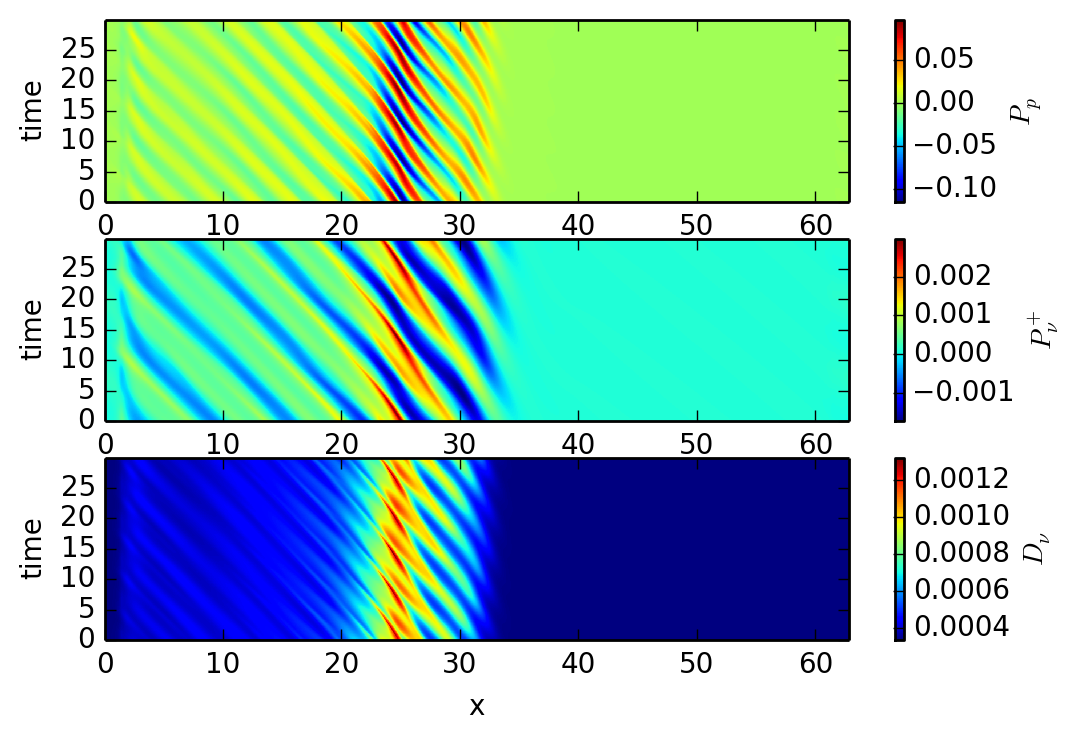}
 \caption{
  $x$-$t$ plot of $P_p$, $P_\nu^+$, and $D_\nu$.
  The characteristic structure around $x\sim 25$ corresponds to the vortex ejection processes.
 }
 \label{fig:xtplot}
\end{figure}

To focus on its streamwise inhomogeneity,
we consider the $y$-averaged energy balance equation:
\begin{equation*}
 \diffp{t}{E} + \dx (J_u + J_\nu) = P_p + P_\nu - D_\nu + F.
 \label{eq:EnergyBalance}
\end{equation*}
It should be noted that
the energy is defined in the interface frame:
$ E(x,t) = \inty \| \vu \|^2 /2$.
Since the walls move, there is an energy injection due to
the viscosity on the walls
$P_\nu = P_\nu^+ + P_\nu^-$,
where $P_\nu^\pm = \mp c_I \partial_y u_x |_{y=\pm 1} / Re $
in addition to the bulk viscous dissipation
\begin{equation*}
 D_\nu = \frac{1}{Re} \inty \left( 2\left( \partial_x u_x \right)^2
                + \left( \partial_x u_y \right)^2
                + \left( \partial_y u_x \right)^2
             \right).
\end{equation*}
The term $ P_p(x,t) = - \inty \left( \vu \cdot \nabla \right) p$
represents the energy injection due to the pressure gradient,
and takes both positive and negative values.
$P_p > 0$ means the flow accelerated by the pressure gradient,
and $P_p < 0$ does the flow against the pressure gradient.
$P_p$ balances almost with the gradient of the energy flux $\partial_x J_u$
and their spatial means are smaller
than those of the viscous terms $P_\nu$ and $D_\nu$.
The flux due to the viscosity $J_\nu$ is negligible, and 
thus neglected hereafter.
$F$ is the energy damping by the filter term.
The three terms $P_p$, $P_\nu^+$, and $D_\nu$ are 
displayed in \cref{fig:xtplot},
which illustrates the traveling of each structures.
Reflecting the chaotic nature of the interface, 
these values are not exactly periodic.
$P_\nu^-(x,t)$ nearly equals to $-P_\nu^+(x,t + T_p/2)$,
where $T_p \sim 15$ denotes an approximate period of 
the recurrent motion at each point,
and thus $P_\nu$ is recurrent with the half 
period $T_p/2$ like as $P_p$ and $D_\nu$.

To confirm that the chaotic interface maintains itself in terms of 
energy balance,
the energy balance equation is averaged over the interface 
region $x \in [20, 34]$:
\begin{equation*}
 \diff{t}{E_I} + \Delta J_u = P_{p, I} + P_{\nu, I} - D_{\nu, I},
\end{equation*}
where the inferior $\cdot_I$ denotes the average over the interface,
and $\Delta J_u(t) = J_u(34,t ) - J_u(20,t)$.
The time average $\left< \cdot \right>$ of these terms are calculated:
$\left< dE_I/dt \right> = -6.7 \times 10^{-6} \approx 0$,
$\left<\Delta J_u \right> = 3.0 \times 10^{-4}$,
$\left<P_{p,I}\right> = 4.3 \times 10^{-4}$,
$\left<P_{\nu,I}\right> =1.2 \times 10^{-3}$, and
$\left<D_{\nu,I}\right> = 1.3 \times 10^{-3}$.
It should be noted that there is the averaged energy 
leak $\left< \Delta J_u \right> > 0$,
which means that the chaotic interface is self-sustainable
in terms of the time-averaged energy balance, and 
even an energy supplier to the weak turbulence.
This energy leak reflects the functionality of the chaotic interface,
i.e. the chaotic interface sustains the weak turbulence.
The right after region of the chaotic interface
has larger energy or stronger turbulent intensity than
the downstream side of the weak turbulent region or asympototic chaotic TS-wave.
This convective relaxation process from this energy excess state
to the asympototic chaotic TS-wave state
is similar to a temporal relaxation process of a minimal 2D channel flow,
which is not shown in this paper.
This similarity and the relationship to the phase reduction scenario
is left to future works.

\section{ejection-jet cycle}
\begin{figure}
 \centering
 \includegraphics[width=\columnwidth]{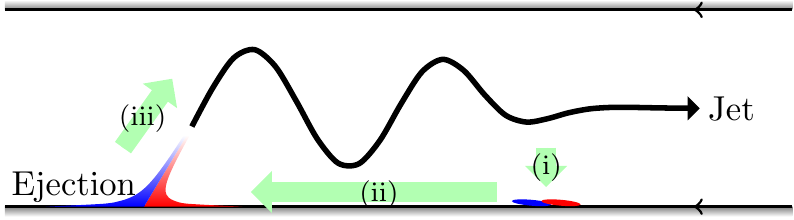}
 \caption{
  A schematic view of the ejection-jet cycle.
 }
 \label{fig:ejection-jet-cycle}
\end{figure}
Here we give a concrete description
of the self-sustaining mechanism of the chaotic interface.
This sustaining process is constituted by the interaction
among vortex ejections on the walls and the meandering jet
in the bulk region.
This collective dynamics is further split into three steps
as summarized in \cref{fig:ejection-jet-cycle}.
In the step (i), a pair of sheet-like vortices is excited
by the instability of the laminar flow near the wall
triggered by the meandering jet.
The amplitude of the meandering decays,
and the jet gets straight as going upstream.
This suggests that a straight jet is convectively stable.
Since the straight jet does not excite the vortex pair,
it does not appear in $x > 35$.

\begin{figure}
 \centering
 \includegraphics[width=\columnwidth]{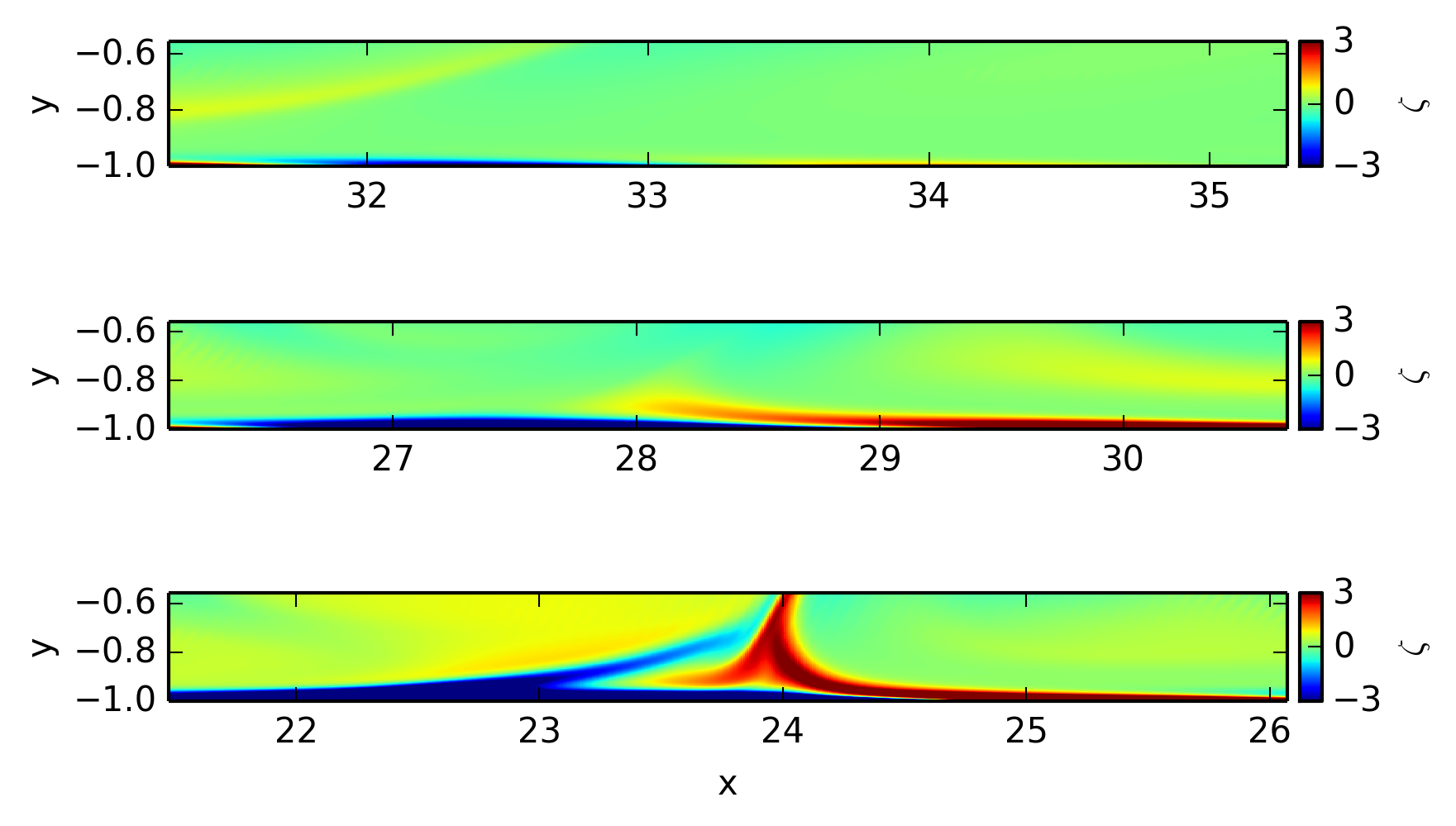}
 \caption{
  The growth of the vortex pair.
  Three figures display different parts of the same snapshot.
  Regions $\zeta > 3$ ($\zeta < -3$) are colored
  by the same color of $\zeta = 3$ ($\zeta = -3$) to emphasize 
  the bulk vorticity,
  though the vorticity field $\zeta(x,y)$ reaches
  its maximum $\zeta_{\rm max} \simeq 25$ around $x \simeq 24$.
 }
 \label{fig:ejection_tl}
\end{figure}

The step (ii) is the convective growth of the vortex pair.
The thin vortex pair generated in the step (i)
grows up into an intense vortex ejection.
This process is displayed in \cref{fig:ejection_tl},
which picks up three continuing parts from a snapshot.
Since the vortex pairs grow convectively,
one snapshot of the entire channel gives three snapshots of the growing vortex pairs.
In the energy viewpoint,
it should be emphasized that this instability is not absolute
but convective in both the interface frame and the laboratory frame.
In the laboratory frame,
the vortex pair goes upstream at a constant speed $c_v \simeq 0.5 > 0$,
and, in the interface frame, it goes downstream at $c_v - c_I \simeq -0.35 < 0$.

The step (iii) is the vortex ejection process,
which excites the jet and makes it meander.
The ejection occurs on the downstream side of the chaotic interface,
namely around $22 < x < 28$.
Then the cycle is closed,
and we call this cycle an  ``ejection-jet'' cycle (EJC).
A very strong shear accompanies this vortex ejection process.
The wall unit $l_\tau$ is estimated at $2.1 \times 10^{-3}$,
and the friction Reynolds number $Re_\tau = l_\tau^{-1}$ is about $460$.
This means that the width of the interface is $5000$ times larger 
than $l_\tau$.
Therefore, we should regard this interface structure
as a large-scale motion in the wall-turbulence context.
After the intensive ejection process, the vortex structures 
are swept downstream,
and this corresponds to the leak of the energy
 $\left< \Delta J_u \right>$ from the interface 
to the weak turbulence region.

To complete the EJC model,
let us consider how the invading speed $c_I$ is determined.
There are two dynamical processes,
the convective growth of the vortex pair
and the decay of the jet meandering.
First, we suppose that both the traveling speed of each vortex pair $c_v$
and the period $T_v$ necessary to grow up are constant.
From \cref{fig:xtplot} we estimate them
at $c_v \simeq 0.5 \pm 0.05$ and $T_v \simeq 20\pm 2$
\footnote{
  The difference between $T_p = 15$ and $T_v = 20$
  is due to the difference in the phase of recurrent motion.
  In other words, the ejection of the grown vortex pair
  and the birth of the next vortex pair are not simultaneous.
}.
Their inaccuracies are due to the inaccurate definitions of them,
and more accurate and quantitative arguments are left to future works.
Then $d(c_I) \defeq | c_v - c_I | T_v$ denotes the distance between
the birth point of the vortex pair and its ejection point.
Next, we introduce a characteristic length $\lambda$
of the decay of the meandering.
Since this process is a nonlinear energy redistribution,
we cannot define it from the spatial linear decay rate,
but instead we measure the distance between the point where
$\left< \max_{y} |u_y(x,y)| \right>$
takes its maximum ($x = 25.4$)
and the point where it becomes almost zero first ($x \simeq 38 \pm 2$).
Since no vortex pair is excited when the jet does not meander,
the EJC model requires these two length are equal:
\begin{equation*}
 |c_v - c_I| T_v = \lambda.
\end{equation*}
This condition connects two values characterizing the different
dynamical processes,
and thus we should regard this condition as a self-consistent 
equation for $c_I$.
The above estimates are consistent with $c_I = 0.855$.

Let us review the EJC model by introducing filtered simulations.
We make other three runs in which the filtered region $\Omega$ is set
to damp one of the specific processes, namely,
(a) weak turbulence, (b) vortex ejection, and (c) vortex pair excitation.
Although we cannot split out each primary dynamics completely,
these filtered simulations help us confirm the EJC model.
These simulations use a snapshot of the previous simulation 
as an initial value,
and animations visualized by the turbulent vorticity are included 
in the supplementary materials.

Case (a): we set $\Omega^{(a)}= [0,22]\times[-1,1]$
to damp the weak turbulent region, and to confirm
the self-sustainability of the chaotic interface.
In this setting we yield a permanent chaotic interface,
whose invading speed and the spatial structure are hardly changed.
We conclude that the following weak turbulence is
additional as assumed in the EJC model.
Furthermore, the selection process of $c_I$ and
the spatial structure  is completely closed in the chaotic interface.
In other words, the weak turbulence region
plays no role in the selection process. 

Case (b): we set $\Omega^{(b)} = [0,30] \times [-1,1]$ to
confirm that the jet is maintained by the acceleration 
due to the vortex ejection.
If the meandering jet is self-sustaining,
this simulation could yield a permanent finite amplitude solution.
However, the laminar flow has occupied whole region.
In this sense, the meandering of the jet is
only a component mechanism of this self-sustaining process, 
and is not self-sustaining.

Case (c): we set $\Omega^{(c)} = [30,20\pi] \times [-1,1]$ 
to obstruct the step (i).
In this case the non-filtered region of the chaotic interface ($20 < x < 30$)
keeps alive on the same position until $t \lesssim 20$,
and then it travels downstream.
This time lag corresponds to the growth time $T_v$ of the vortex ejection,
and thus this result also supports the EJC model.
After a long transient,
another chaotic interface is reconstructed around $15\lesssim x \lesssim 27$,
and their invading speed and spatial structure are same as the previous one.
This result insists that the chaotic interface structure is robust
while there is a laminar flow on its upstream.
This robustness is an important issue for the pattern selection problem,
but the current framework of the dynamical systems approach lacks 
tools applicable for settling the issue.

\section{Concluding Remarks}
We have investigated the self-sustainability and functionality
of CI in two-dimensional channel flow as an example of FCS,
which yields the inhomogeneity between two asymptotic homogeneous states,
upstream laminar flow and downstream chaotic TS-wave.
We have introduced a phenomenology summarized in the EJC model,
which consists of the vortex ejection and the meandering jet.
The localized dynamics of CI is isolated by the filtered simulation,
and deconstructed by the energy balance analysis.
The EJC model well represents both the invading process on its front
and its functionality in sustaining the weak turbulence on its tail.
As a result, however, this functionality prevents us 
from obtaining an exact localized solution corresponding 
to the chaotic interface as done for various coherent structures
because weak turbulence must attach to the interface.
The damping filter works effectively
in isolating the localized dynamics of CI.

The self-sustaining mechanism described by the EJC model
is also an example for collective dynamics of multiscale structures.
Different from Waleffe's self-sustaining process \cite{CambridgeJournals:353953}
which utilizes an absolute instability, the EJC model does
a convective instability, which needs a sufficient space to grow up.
The convective instability makes it possible
for the structures of different scales to interact with each other,
namely the meandering jet of large scale and the wall shear of small scale.
This multiscale interaction mechanism may
be applied for the large-scale motion in three-dimensional wall-turbulence \cite{Toh2005},
although the chaotic nature of CI is
far weaker than that of three-dimensional wall-turbulence.
Furthermore, it may also be a prototype
for more general multiscale collective dynamics.

We have introduced the functional coherent structure (FCS),
which extends the well-known coherent structure perspective.
Previous studies have focused on the self-sustainability of coherent structures,
but we do on its additional functionality.
We expect that the idea to assign functionalities of turbulence
to localized coherent structures may work well for other cases.
Energy and momentum transfers in fully-developed wall-turbulence
are possible applications since functional Waleffe's SSP, if it exists,
may be embedded near the wall.
For further development of the building block strategy,
we will have to combine this functional building block scenario
with the phase reduction scenario.
In other words, we have to establish a framework involving
phenomenological low-dimensional models of FCS and their interactions,
and it is left to future works.
This framework will be an essential tool for the dynamical systems approach
to inhomogeneous turbulence and more general spatiotemporal chaotic systems.

\begin{acknowledgments}
 This work is supported by the Grant-in-Aid for JSPS Fellows No. 26$\cdot$1005
 and the Grants for Excellent Graduate Schools
 ``The Next Generation of Physics, Spun from Universality and Emergence''
 from the Ministry of Education, Culture, Sports, Science, and Technology (MEXT) of Japan,
 and also partially by JSPS KAKENHI Grant Number 22540386.
 A part of numerical calculations were carried out on SR16000 at YITP in Kyoto University.
\end{acknowledgments}

\bibliographystyle{apsrev4-1}
\bibliography{library}

\begin{thebibliography}{20}%
\makeatletter
\providecommand \@ifxundefined [1]{%
 \@ifx{#1\undefined}
}%
\providecommand \@ifnum [1]{%
 \ifnum #1\expandafter \@firstoftwo
 \else \expandafter \@secondoftwo
 \fi
}%
\providecommand \@ifx [1]{%
 \ifx #1\expandafter \@firstoftwo
 \else \expandafter \@secondoftwo
 \fi
}%
\providecommand \natexlab [1]{#1}%
\providecommand \enquote  [1]{``#1''}%
\providecommand \bibnamefont  [1]{#1}%
\providecommand \bibfnamefont [1]{#1}%
\providecommand \citenamefont [1]{#1}%
\providecommand \href@noop [0]{\@secondoftwo}%
\providecommand \href [0]{\begingroup \@sanitize@url \@href}%
\providecommand \@href[1]{\@@startlink{#1}\@@href}%
\providecommand \@@href[1]{\endgroup#1\@@endlink}%
\providecommand \@sanitize@url [0]{\catcode `\\12\catcode `\$12\catcode
  `\&12\catcode `\#12\catcode `\^12\catcode `\_12\catcode `\%12\relax}%
\providecommand \@@startlink[1]{}%
\providecommand \@@endlink[0]{}%
\providecommand \url  [0]{\begingroup\@sanitize@url \@url }%
\providecommand \@url [1]{\endgroup\@href {#1}{\urlprefix }}%
\providecommand \urlprefix  [0]{URL }%
\providecommand \Eprint [0]{\href }%
\providecommand \doibase [0]{http://dx.doi.org/}%
\providecommand \selectlanguage [0]{\@gobble}%
\providecommand \bibinfo  [0]{\@secondoftwo}%
\providecommand \bibfield  [0]{\@secondoftwo}%
\providecommand \translation [1]{[#1]}%
\providecommand \BibitemOpen [0]{}%
\providecommand \bibitemStop [0]{}%
\providecommand \bibitemNoStop [0]{.\EOS\space}%
\providecommand \EOS [0]{\spacefactor3000\relax}%
\providecommand \BibitemShut  [1]{\csname bibitem#1\endcsname}%
\let\auto@bib@innerbib\@empty
%</preamble>
\bibitem [{\citenamefont {Venugopalan}(2014)}]{Venugopalan2014}%
  \BibitemOpen
  \bibfield  {author} {\bibinfo {author} {\bibfnamefont {R.}~\bibnamefont
  {Venugopalan}},\ }\href {\doibase 10.1016/j.nuclphysa.2014.04.030} {\bibfield
   {journal} {\bibinfo  {journal} {Nuclear Physics A}\ }\textbf {\bibinfo
  {volume} {928}},\ \bibinfo {pages} {209} (\bibinfo {year} {2014})},\ \Eprint
  {http://arxiv.org/abs/1404.6976} {arXiv:1404.6976} \BibitemShut {NoStop}%
\bibitem [{\citenamefont {Balbus}\ and\ \citenamefont
  {Hawley}(1998)}]{Balbus1998}%
  \BibitemOpen
  \bibfield  {author} {\bibinfo {author} {\bibfnamefont {S.}~\bibnamefont
  {Balbus}}\ and\ \bibinfo {author} {\bibfnamefont {J.}~\bibnamefont
  {Hawley}},\ }\href {\doibase 10.1103/RevModPhys.70.1} {\bibfield  {journal}
  {\bibinfo  {journal} {Reviews of Modern Physics}\ }\textbf {\bibinfo {volume}
  {70}},\ \bibinfo {pages} {1} (\bibinfo {year} {1998})}\BibitemShut {NoStop}%
\bibitem [{\citenamefont {Frisch}(1995)}]{Frisch1995}%
  \BibitemOpen
  \bibfield  {author} {\bibinfo {author} {\bibfnamefont {U.}~\bibnamefont
  {Frisch}},\ }\href@noop {} {\emph {\bibinfo {title} {{Turbulence: the legacy
  of AN Kolmogorov}}}}\ (\bibinfo  {publisher} {Cambridge university press},\
  \bibinfo {year} {1995})\BibitemShut {NoStop}%
\bibitem [{\citenamefont {Kawahara}\ \emph {et~al.}(2011)\citenamefont
  {Kawahara}, \citenamefont {Uhlmann},\ and\ \citenamefont {van
  Veen}}]{Kawahara2012}%
  \BibitemOpen
  \bibfield  {author} {\bibinfo {author} {\bibfnamefont {G.}~\bibnamefont
  {Kawahara}}, \bibinfo {author} {\bibfnamefont {M.}~\bibnamefont {Uhlmann}}, \
  and\ \bibinfo {author} {\bibfnamefont {L.}~\bibnamefont {van Veen}},\ }\href
  {\doibase 10.1146/annurev-fluid-120710-101228} {\bibfield  {journal}
  {\bibinfo  {journal} {Annual Review of Fluid Mechanics}\ }\textbf {\bibinfo
  {volume} {44}},\ \bibinfo {pages} {203} (\bibinfo {year} {2011})},\ \Eprint
  {http://arxiv.org/abs/1108.0975} {arXiv:1108.0975} \BibitemShut {NoStop}%
\bibitem [{\citenamefont {Jimeez}\ and\ \citenamefont
  {Moin}(1991)}]{Jimenez1991}%
  \BibitemOpen
  \bibfield  {author} {\bibinfo {author} {\bibfnamefont {J.}~\bibnamefont
  {Jimeez}}\ and\ \bibinfo {author} {\bibfnamefont {P.}~\bibnamefont {Moin}},\
  }\href {\doibase 10.1017/S0022112091002033} {\bibfield  {journal} {\bibinfo
  {journal} {Journal of Fluid Mechanics}\ }\textbf {\bibinfo {volume} {225}},\
  \bibinfo {pages} {213} (\bibinfo {year} {1991})}\BibitemShut {NoStop}%
\bibitem [{\citenamefont {Avila}\ \emph {et~al.}(2013)\citenamefont {Avila},
  \citenamefont {Mellibovsky}, \citenamefont {Roland},\ and\ \citenamefont
  {Hof}}]{Avila2013}%
  \BibitemOpen
  \bibfield  {author} {\bibinfo {author} {\bibfnamefont {M.}~\bibnamefont
  {Avila}}, \bibinfo {author} {\bibfnamefont {F.}~\bibnamefont {Mellibovsky}},
  \bibinfo {author} {\bibfnamefont {N.}~\bibnamefont {Roland}}, \ and\ \bibinfo
  {author} {\bibfnamefont {B.}~\bibnamefont {Hof}},\ }\href {\doibase
  10.1103/PhysRevLett.110.224502} {\bibfield  {journal} {\bibinfo  {journal}
  {Physical Review Letters}\ }\textbf {\bibinfo {volume} {110}},\ \bibinfo
  {pages} {224502} (\bibinfo {year} {2013})},\ \Eprint
  {http://arxiv.org/abs/1212.0230v3} {arXiv:1212.0230v3} \BibitemShut {NoStop}%
\bibitem [{\citenamefont {Schneider}\ \emph {et~al.}(2009)\citenamefont
  {Schneider}, \citenamefont {Marinc},\ and\ \citenamefont
  {Eckhardt}}]{CambridgeJournals:7324528}%
  \BibitemOpen
  \bibfield  {author} {\bibinfo {author} {\bibfnamefont {T.}~\bibnamefont
  {Schneider}}, \bibinfo {author} {\bibfnamefont {D.}~\bibnamefont {Marinc}}, \
  and\ \bibinfo {author} {\bibfnamefont {B.}~\bibnamefont {Eckhardt}},\ }\href
  {\doibase 10.1017/S0022112009993144} {\bibfield  {journal} {\bibinfo
  {journal} {Journal of Fluid Mechanics}\ }\textbf {\bibinfo {volume} {646}},\
  \bibinfo {pages} {15} (\bibinfo {year} {2009})},\ \Eprint
  {http://arxiv.org/abs/0909.0530} {arXiv:0909.0530} \BibitemShut {NoStop}%
\bibitem [{\citenamefont {Schneider}\ \emph {et~al.}(2010)\citenamefont
  {Schneider}, \citenamefont {Gibson},\ and\ \citenamefont
  {Burke}}]{Schneider2010a}%
  \BibitemOpen
  \bibfield  {author} {\bibinfo {author} {\bibfnamefont {T.~M.}\ \bibnamefont
  {Schneider}}, \bibinfo {author} {\bibfnamefont {J.~F.}\ \bibnamefont
  {Gibson}}, \ and\ \bibinfo {author} {\bibfnamefont {J.}~\bibnamefont
  {Burke}},\ }\href {\doibase 10.1103/PhysRevLett.104.104501} {\bibfield
  {journal} {\bibinfo  {journal} {Physical Review Letters}\ }\textbf {\bibinfo
  {volume} {104}},\ \bibinfo {pages} {104501} (\bibinfo {year} {2010})},\
  \Eprint {http://arxiv.org/abs/0912.2739} {arXiv:0912.2739} \BibitemShut
  {NoStop}%
\bibitem [{\citenamefont {Eckhardt}(2014)}]{Eckhardt2014a}%
  \BibitemOpen
  \bibfield  {author} {\bibinfo {author} {\bibfnamefont {B.}~\bibnamefont
  {Eckhardt}},\ }\href {\doibase 10.1017/jfm.2014.442} {\bibfield  {journal}
  {\bibinfo  {journal} {Journal of Fluid Mechanics}\ }\textbf {\bibinfo
  {volume} {758}},\ \bibinfo {pages} {1} (\bibinfo {year} {2014})}\BibitemShut
  {NoStop}%
\bibitem [{\citenamefont {Khapko}\ \emph {et~al.}(2013)\citenamefont {Khapko},
  \citenamefont {Kreilos}, \citenamefont {Schlatter}, \citenamefont {Duguet},
  \citenamefont {Eckhardt},\ and\ \citenamefont {Henningson}}]{Khapko2013}%
  \BibitemOpen
  \bibfield  {author} {\bibinfo {author} {\bibfnamefont {T.}~\bibnamefont
  {Khapko}}, \bibinfo {author} {\bibfnamefont {T.}~\bibnamefont {Kreilos}},
  \bibinfo {author} {\bibfnamefont {P.}~\bibnamefont {Schlatter}}, \bibinfo
  {author} {\bibfnamefont {Y.}~\bibnamefont {Duguet}}, \bibinfo {author}
  {\bibfnamefont {B.}~\bibnamefont {Eckhardt}}, \ and\ \bibinfo {author}
  {\bibfnamefont {D.~S.}\ \bibnamefont {Henningson}},\ }\href {\doibase
  10.1017/jfm.2013.20} {\bibfield  {journal} {\bibinfo  {journal} {Journal of
  Fluid Mechanics}\ }\textbf {\bibinfo {volume} {717}},\ \bibinfo {pages} {R6}
  (\bibinfo {year} {2013})}\BibitemShut {NoStop}%
\bibitem [{\citenamefont {Kuramoto}(2003)}]{Kuramoto2003}%
  \BibitemOpen
  \bibfield  {author} {\bibinfo {author} {\bibfnamefont {Y.}~\bibnamefont
  {Kuramoto}},\ }\href@noop {} {\emph {\bibinfo {title} {{Chemical
  oscillations, waves, and turbulence}}}}\ (\bibinfo  {publisher} {Courier
  Corporation},\ \bibinfo {year} {2003})\BibitemShut {NoStop}%
\bibitem [{\citenamefont {Barkley}(2011)}]{Barkley2011}%
  \BibitemOpen
  \bibfield  {author} {\bibinfo {author} {\bibfnamefont {D.}~\bibnamefont
  {Barkley}},\ }\href {\doibase 10.1103/PhysRevE.84.016309} {\bibfield
  {journal} {\bibinfo  {journal} {Physical Review E - Statistical, Nonlinear,
  and Soft Matter Physics}\ }\textbf {\bibinfo {volume} {84}},\ \bibinfo
  {pages} {016309} (\bibinfo {year} {2011})},\ \Eprint
  {http://arxiv.org/abs/1101.4125} {arXiv:1101.4125} \BibitemShut {NoStop}%
\bibitem [{\citenamefont {Jim\'{e}nez}(1990)}]{Jimenez1990}%
  \BibitemOpen
  \bibfield  {author} {\bibinfo {author} {\bibfnamefont {J.}~\bibnamefont
  {Jim\'{e}nez}},\ }\href {\doibase 10.1017/S0022112090001008} {\enquote
  {\bibinfo {title} {{Transition to turbulence in two-dimensional Poiseuille
  flow}},}\ } (\bibinfo {year} {1990})\BibitemShut {NoStop}%
\bibitem [{\citenamefont {Fortin}\ \emph {et~al.}(1994)\citenamefont {Fortin},
  \citenamefont {Jardak}, \citenamefont {Gervais},\ and\ \citenamefont
  {Pierre}}]{Fortin1994a}%
  \BibitemOpen
  \bibfield  {author} {\bibinfo {author} {\bibfnamefont {A.}~\bibnamefont
  {Fortin}}, \bibinfo {author} {\bibfnamefont {M.}~\bibnamefont {Jardak}},
  \bibinfo {author} {\bibfnamefont {J.}~\bibnamefont {Gervais}}, \ and\
  \bibinfo {author} {\bibfnamefont {R.}~\bibnamefont {Pierre}},\ }\href
  {\doibase 10.1006/jcph.1994.1210} {\bibfield  {journal} {\bibinfo  {journal}
  {Journal of Computational Physics}\ }\textbf {\bibinfo {volume} {115}},\
  \bibinfo {pages} {455} (\bibinfo {year} {1994})}\BibitemShut {NoStop}%
\bibitem [{\citenamefont {Umeki}(1994)}]{Umeki1994}%
  \BibitemOpen
  \bibfield  {author} {\bibinfo {author} {\bibfnamefont {M.}~\bibnamefont
  {Umeki}},\ }\href {\doibase 10.1016/0169-5983(94)90007-8} {\enquote {\bibinfo
  {title} {{Numerical simulation of plane Poiseuille turbulence}},}\ }
  (\bibinfo {year} {1994})\BibitemShut {NoStop}%
\bibitem [{\citenamefont {Rauh}\ \emph {et~al.}(1995)\citenamefont {Rauh},
  \citenamefont {Zachrau},\ and\ \citenamefont {Zoller}}]{Rauh1995}%
  \BibitemOpen
  \bibfield  {author} {\bibinfo {author} {\bibfnamefont {a.}~\bibnamefont
  {Rauh}}, \bibinfo {author} {\bibfnamefont {T.}~\bibnamefont {Zachrau}}, \
  and\ \bibinfo {author} {\bibfnamefont {J.}~\bibnamefont {Zoller}},\ }\href
  {\doibase 10.1016/0167-2789(95)00194-9} {\enquote {\bibinfo {title}
  {{Nonlinear stability analysis of plane poiseuille flow by normal forms}},}\
  } (\bibinfo {year} {1995})\BibitemShut {NoStop}%
\bibitem [{\citenamefont {Teramura}\ and\ \citenamefont
  {Toh}(2014)}]{Teramura2014}%
  \BibitemOpen
  \bibfield  {author} {\bibinfo {author} {\bibfnamefont {T.}~\bibnamefont
  {Teramura}}\ and\ \bibinfo {author} {\bibfnamefont {S.}~\bibnamefont {Toh}},\
  }\href {\doibase 10.1103/PhysRevE.89.052910} {\bibfield  {journal} {\bibinfo
  {journal} {Physical Review E - Statistical, Nonlinear, and Soft Matter
  Physics}\ }\textbf {\bibinfo {volume} {89}},\ \bibinfo {pages} {052910}
  (\bibinfo {year} {2014})}\BibitemShut {NoStop}%
\bibitem [{Note1()}]{Note1}%
  \BibitemOpen
  \bibinfo {note} {The difference between $T_p = 15$ and $T_v = 20$ is due to
  the difference in the phase of recurrent motion. In other words, the ejection
  of the grown vortex pair and the birth of the next vortex pair are not
  simultaneous.}\BibitemShut {Stop}%
\bibitem [{\citenamefont {Hamilton}\ \emph {et~al.}(1995)\citenamefont
  {Hamilton}, \citenamefont {Kim},\ and\ \citenamefont
  {Waleffe}}]{CambridgeJournals:353953}%
  \BibitemOpen
  \bibfield  {author} {\bibinfo {author} {\bibfnamefont {J.~M.}\ \bibnamefont
  {Hamilton}}, \bibinfo {author} {\bibfnamefont {J.}~\bibnamefont {Kim}}, \
  and\ \bibinfo {author} {\bibfnamefont {F.}~\bibnamefont {Waleffe}},\ }\href
  {\doibase 10.1017/S0022112095000978} {\enquote {\bibinfo {title}
  {{Regeneration mechanisms of near-wall turbulence structures}},}\ } (\bibinfo
  {year} {1995})\BibitemShut {NoStop}%
\bibitem [{\citenamefont {TOH}\ and\ \citenamefont {ITANO}(2005)}]{Toh2005}%
  \BibitemOpen
  \bibfield  {author} {\bibinfo {author} {\bibfnamefont {S.}~\bibnamefont
  {TOH}}\ and\ \bibinfo {author} {\bibfnamefont {T.}~\bibnamefont {ITANO}},\
  }\href {\doibase 10.1017/S002211200400237X} {\enquote {\bibinfo {title}
  {{Interaction between a large-scale structure and near-wall structures in
  channel flow}},}\ } (\bibinfo {year} {2005})\BibitemShut {NoStop}%
\end{thebibliography}%
\end{document}